\documentclass[12pt,english,amsmath,amssymb,aps,pra,letterpaper]{revtex4-1}
\usepackage{graphicx}
\usepackage[T1]{fontenc}
\usepackage[latin9]{inputenc}
\usepackage[letterpaper]{geometry}
\usepackage{units}
\usepackage{esint}
\usepackage{braket}

\makeatletter

\usepackage{mathrsfs}
\usepackage{bbm}
\usepackage{bm}

\makeatother

\usepackage{babel}
\usepackage{float}

\begin{document}

\title{Probing the Role of the Eighth Bacteriochlorophyll in \textit{holo}-FMO Complex by Simulated Two-Dimensional Electronic Spectroscopy}

\author{Shu-Hao Yeh}
\affiliation{Department of Chemistry and Birck Nanotechnology Center, Purdue University,
West Lafayette, IN 47907, USA}
\author{Sabre Kais}
\thanks{Corresponding author, kais@purdue.edu}
\affiliation{Department of Chemistry and Birck Nanotechnology Center, Purdue University,
West Lafayette, IN 47907, USA}
\affiliation{Qatar Environment and Energy Research Institute, Qatar Foundation,
Doha, Qatar}

\begin{abstract}
The Fenna-Matthews-Olson (FMO) protein-pigment complex acts as a molecular wire between the outer antenna system and the reaction center (RC); it is an important model system to study the excitonic energy transfer. Recent crystallographic studies report the existence of an additional (eighth) bacteriochlorophyll a (BChl a). To understand the functionality of this eighth BChl, we simulated the two-dimensional electronic spectra of both the 7-site (\textit{apo} form) and the 8-site (\textit{holo} form) variant of the FMO complex from green sulfur bacteria, \textit{Prosthecochloris aestuarii}. By comparing the difference between the spectrum, it was found that the eighth BChl can affect two different excitonic energy transfer pathways, these being: (1) directly involve in the first pathway 6 $\rightarrow$ 3 $\rightarrow$ 1 of the \textit{apo} form model by passing the excitonic energy to exciton 6; and (2) increase the excitonic wave function overlap between excitons 4 and 5 in the second pathway (7 $\rightarrow$ 4,5 $\rightarrow$ 2 $\rightarrow$ 1) and thus increase the possible downward sampling routes across the BChls.
\end{abstract}
\maketitle

\section{Introduction}

Solar energy is one of the most abundant energy source which can be utilized by living organisms. In plants and bacteria, the mechanism that converts this solar electromagnetic energy into chemical energy is called photosynthesis \citep{vanGrondelleBook2000, MolMechPhoto, Renger2001}. One of the most studied model systems is the Fenna-Matthews-Olson (FMO) protein-pigment complex in green sulfur bacteria \cite{FENNA1975, SYBESMA1963, Ishizaki2010}, which funnels the excitonic energy absorbed from the outer antenna system (chlorosome/baseplate) to the reaction center. Excitonic energy transfer (EET) in such photosynthetic protein-pigment complexes has been studied by various nonlinear optical spectroscopy methods such as pump-probe, hole burning, and photon-echo have been applied \citep{vanGrondelleBook2000, Grondelle2006}. Recent experimental studies using 2D electronic photon-echo spectroscopy reported beating patterns with oscillation periods that approximately correspond to the difference of the eigenenergies. This phenomenon has been found in several photosynthetic protein-pigment complexes at both cryogenic temperatures \citep{Engel-2007, Fleming-Science-2007, Calhoun2009} and room temperature \citep{Panitchayangkoon2010, Collini-2010}, which implies that quantum coherence may be a crucial component to the reported high-efficiency energy transfer  \citep{Alan-2008, Cao2009, Caruso2010}.

The FMO complex has a homotrimer structure and in recent crystallographic studies it was found that the presence of an additional (eighth) bacteriochlorophyll \textit{a} (BChl \textit{a}) per subunit, which sit in in the cleft at the protein surface close to the baseplate \citep{Ben-Shem2004, Tronrud2009b}. Schmidt am Busch \textit{et al}. calculated the site energies of BChls in the FMO complex based on a new crystal structure from the green sulfur bacteria,  \textit{Prosthecochloris aestuarii} \citep{Busch2011}. In their calculation it was found that BChl 8 has the highest site energy, and after modelling the baseplate and the reaction center (RC) by two low dielectric constant layers ($\epsilon = 4$) they also found that BChl 8 was red-shifted and the overlap with the baseplate fluorescence was increased. Furthermore, due to the greater vicinity of BChl 8 to the baseplate, BChl 8 being closer than any other BChl of the FMO, it has also been proposed that BChl 8 functions as a linker between the baseplate and the FMO complex. 

Experimental findings of quantum coherence within the photosynthetic protein-pigment complexes also instigated great interest in the use of theoretical models to describe photosynthetic EET. Theoretical description is particularly challenging in some specific systems such as the FMO complex and the B850 ring of the light harvesting complex II (LH2) found in purple bacteria; these difficulties are due to the fact that the magnitude of the electronic coupling and the exciton-phonon coupling are of same order and therefore neither of them can be treated perturbatively. To tackle this problem, a methodology referred to as hierarchical equation of motion (HEOM) has been developed by Tanimura and co-workers \citep{Tanimura-Kubo-1989, Tanimura}; a high temperature approximation (HTA) of HEOM has also been developed by Ishizaki and co-workers to describe the EET dynamics in a model dimer \citep{Ishizaki2009, Fleming-JCP-2009}. The HEOM has already been implemented in calculating the excitonic transfer dynamics of photosynthetic protein-pigment complexes \citep{Strumpfer2009a, Strumpfer2011, Zhu2011, Zhu2012, Yeh2012} and more recently it is also used within simulations of the linear and 2D electronic spectra of the above mentioned systems \citep{Chen2010, Jing2013, Hein2012a}. 

The aim of this study is to understand the role of the BChl 8 in the EET pathway of the FMO complex. To achieve this, we first analyse the difference of excitonic wave function overlap among BChls between the \textit{apo} (7-sites) and \textit{holo} (8-sites) forms and propose the corresponding EET pathways. Then the HEOM is applied to simulate the 2D photon-echo spectra of both forms by using a model Hamiltonian formed in a previous study (PDBID: 3EOJ) \cite{Tronrud2009a}. Meanwhile we have also explored the effect of both temperature and of the the presence of static disorder to the EET.

\newpage


\subsection{Model Hamiltonian for 2D spectra calculation}

The total Hamiltonian for 2D spectra calculation can be described as:
\begin{equation}
H=H_{S}+H_{B}+H_{SB}, \label{eq:Htot}
\end{equation}
where $H_{S}$, $H_{B}$ and $H_{SB}$ represent the Hamiltonian of the system, environment and system-environment coupling respectively. For the system Hamiltonian, $H_{S}$, both the 7-sites and 8-sites single-exciton Hamiltonian of the FMO BChls are obtained from a structural-based theoretical study by Schmidt am Busch et al. \cite{Busch2011}. This single exciton Hamiltonian is given as:
\begin{equation}
H_{1ex}=\sum_{j=1}^{N}\left(\left(\varepsilon_{j}+\lambda_j\right)\ket{j}\bra{j}+\sum_{k\neq j}J_{jk}\left(\ket{j}\bra{k}+\ket{k}\bra{j}\right)\right), \label{eq:H1ex}
\end{equation}
where the state $\ket{j}$ represents the $Q_\textrm{Y}$ excitation of the $j$th BChl with site energy $\varepsilon_{j}$, and $J_{jk}$ corresponds to the excitonic dipole-dipole coupling between BChl $j$ and $k$. $\lambda_j$ is the reorganization energy which is introduced to counteract the energy shift due to the system-bath coupling. It is assumed that this reorganization energy is identical for each site, therefore $\lambda_j = \lambda$.

In order to calculate the two-dimensional electronic spectra it is necessary to include both the ground state and the two-exciton manifold into consideration. In the two-exciton manifold, the system is capable of exciting two different BChls simultaneously and its Hamiltonian can be constructed based on the information of the single-exciton manifold:
\begin{equation}
\begin{split}
\bra{jk}H_{2ex}\ket{j'k'}&=\delta_{jj'}\delta_{kk'}(\varepsilon_{j}+\varepsilon_{k}) + \delta_{jj'}(1-\delta_{kk'})J_{kk'} 
       +\delta_{jk'}(1-\delta_{kj'})J_{kj'} \\
       &+ \delta_{kj'}(1-\delta_{jk'})J_{jk'} +\delta_{kk'}(1-\delta_{jj'})J_{jj'}\;. \label{eq:H2ex}
\end{split}
\end{equation}
$\ket{jk}$ represents a two-exciton state describing the simultaneous excitation of sites $j$ and $k$, where for counting $1 \leq j < k \leq N$. The first term defines the diagonal elements by summing the site energy of the two involving sites. The remaining terms describe the off-diagonal elements by stipulating that the excitonic transition is permitted if the initial and the final states share only one common site, where the strength defined by the coupling of the non-sharing pairs. The final system Hamiltonian used for 2D spectra calculation is in a block diagonal form and can be written as:
\begin{equation}
H_{S} = 
\begin{pmatrix}
0 & 0 & 0 \\
0 & H_{1ex} & 0 \\
0 & 0 & H_{2ex} 
\end{pmatrix} 
\;, \label{eq:Hs}
\end{equation}
where the energy of the ground state is taken to be zero. 

The environment Hamiltonian, $H_{B}$, describes the nuclear degrees of freedom by a phonon bath:
\begin{equation}
H_{B} =\sum_{j=1}^{N}\sum_{\xi}\frac{\hat{p}_{j\xi}^{2}}{2m_{j\xi}}+\frac{1}{2}m_{j\xi}\omega_{j\xi}^{2}\hat{x}_{j\xi}^{2}\;, \label{eq:HB}
\end{equation}
where $m_{j\xi}$, $\omega_{j\xi}$, $\hat{x}_{j\xi}$, $\hat{p}_{j\xi}$ are mass, frequency, position and momentum operator of the $\xi$-th bath oscillator associate with the $j$-th BChl respectively. For the system-environment coupling Hamiltonian, $H_{SB}$, it is assumed that each BChl is linearly coupled to the vibrational modes of its own bath and therefore induces site energy fluctuations independently. This interaction Hamiltonian is given by:

\begin{equation}
H_{SB} =\sum_{j=1}^{N}\sum_{\xi}c_{j\xi}\ket{j}\bra{j}\hat{x}_{j\xi} = \sum_{j=1}^{N}\hat{P}_{j}\hat{B}_{j}, \label{eq:HSB}\\
\end{equation}

where $\hat{P}_{j}=\ket{j}\bra{j}$ is the excitonic projection operator and $\hat{B}_{j}=\sum_{\xi}c_{j\xi} \hat{x}_{j\xi}$ is defined as the collective bath operator. $c_{j\xi}$ represents the electron-phonon coupling between the $j$-th BChl and the $\xi$-th phonon mode.

\subsection{Optical response functions}
The interaction between electric dipoles and a weak electromagnetic field can be simulated by using optical response functions. These response functions describe how a weakly-perturbed nonequilibrium system deviates from its equilibrium behaviors. The 2D spectra can be simulated by first calculating the third-order optical response function \cite{MukamelBook1}, which takes the form:
\begin{equation}
R^{(3)}(t_3,t_2,t_1)= -\frac{i}{\hbar^3}  \left\langle\left[\left[\left[ \hat{\mu}(t_1+t_2+t_3),\hat{\mu}(t_1+t_2)\right] ,\hat{\mu}(t_1)\right], \hat{\mu}\right]\right\rangle\;. \label{eq:3rdRF}\\
\end{equation}
Where $\hat{\mu}(t) \equiv e^{iHt/\hbar}\hat{\mu}e^{-iHt/\hbar}$ is the dipole operator within Heisenberg picture and $\langle \cdots \rangle \equiv \mathrm{Tr}\left\{\cdots\hat{\rho}_{eq}\right\}$. Because the excitation energy is much larger than the scale of the thermal energy ($k_{\mathrm{B}}T$) for our system of interest, it can be assumed that the system is initially equilibrated in its ground state. The initial equilibrium state can thus be written as $\hat{\rho}_{eq} = \ket{0}\bra{0} \otimes e^{-\beta H_B}/\mathrm{Tr}_B\left\{e^{-\beta H_B}\right\}$, where $\beta$ refrains its standard thermodynamic definition as $1/k_{\mathrm{B}}T$ and $\ket{0}$ represents the ground state of the system. 
$\hat{\mu}$ is the total transition dipole operator defined as $\hat{\mu} = \hat{\mu}_- + \hat{\mu}_+$, where $\hat{\mu}_+ = \hat{\mu}_-^{\dag}$ and
\begin{equation}
\hat{\mu}_- = \sum_{j=1}^{N} \mu_j\ket{0}\bra{j} + \sum_{j=1}^{N-1}\sum_{k=j+1}^{N} \mu_j\ket{k}\bra{jk} +\mu_k\ket{j}\bra{jk}. \label{eq:op_dipole}\\
\end{equation}

In 2D photon echo experiment, three laser pulses with wave vectors $\bm{k}_1$, $\bm{k}_2$, and $\bm{k}_3$ are sequentially applied to the analyte so to generate a third-order polarization, and this nonlinear response can be detected along a phase-matched direction \citep{Khalil2003, Cho2001, Cho2008, Abramavicius2009}.  The rephasing signal is detected along the direction $\bm{k}_\mathrm{RP}=-\bm{k}_1+\bm{k}_2+\bm{k}_3$ and its contribution in response function can be represented as:
\begin{equation}
R_\mathrm{RP}^{(3)}(t_3,t_2,t_1)= -\frac{i}{\hbar^3} \mathrm{Tr}\left\{ \hat{\mu}_- \hat{\mathcal{G}}(t_3)\hat{\mu}_+^{\times} \hat{\mathcal{G}}(t_2)\hat{\mu}_+^{\times} \hat{\mathcal{G}}(t_1)\hat{\mu}_-^{\times}\hat{\rho}_{eq}\right\} , \label{eq:3rdRP}\\
\end{equation}
with using the superoperator notation $\hat{\mathcal{O}}^{\times} \hat{f} \equiv \hat{\mathcal{O}}\hat{f} - \hat{f}\hat{\mathcal{O}}$ for any operator $\hat{\mathcal{O}}$ and operand operator $\hat{f}$. The nonrephasing signal is detected along the direction $\bm{k}_\mathrm{NR}=\bm{k}_1-\bm{k}_2+\bm{k}_3$ and its contribution in response function can be written as:
\begin{equation}
R_\mathrm{NR}^{(3)}(t_3,t_2,t_1)= -\frac{i}{\hbar^3} \mathrm{Tr}\left\{ \hat{\mu}_- \hat{\mathcal{G}}(t_3)\hat{\mu}_+^{\times} \hat{\mathcal{G}}(t_2)\hat{\mu}_-^{\times} \hat{\mathcal{G}}(t_1)\hat{\mu}_+^{\times}\hat{\rho}_{eq}\right\}. \label{eq:3rdNR}\\
\end{equation}
By carrying out a double Fourier transform of Eq. \eqref{eq:3rdRP} and Eq. \eqref{eq:3rdNR} for temporal variables $t_1$ and $t_3$, the absorptive part of the 2D electronic spectra then can be calculated as $S(\omega_3, t_2, \omega_1) \equiv S_\mathrm{RP}(\omega_3, t_2, \omega_1) + S_\mathrm{NR}(\omega_3, t_2, \omega_1)$, where the rephasing contribution is:
\begin{equation}
S_\mathrm{RP}^{(3)}(\omega_3,t_2,\omega_1) \equiv \mathrm{Im}\int_0^{\infty}dt_1 \int_0^{\infty} dt_3 e^{i(\omega_1 t_1+\omega_3 t_3)} R_\mathrm{RP}^{(3)}(t_3,t_2,t_1) \;, \label{eq:3rdSRP}\\
\end{equation}
and the nonrephasing contribution is:
\begin{equation}
S_\mathrm{NR}^{(3)}(\omega_3,t_2,\omega_1) \equiv \mathrm{Im}\int_0^{\infty}dt_1 \int_0^{\infty} dt_3 e^{i(-\omega_1 t_1+\omega_3 t_3)} R_\mathrm{NR}^{(3)}(t_3,t_2,t_1) \;. \label{eq:3rdSNR}\\
\end{equation}

Because the third-order nonlinear response function possesses inversion symmetry along the of laser polarization, the rotational average of the 2D spectra is calculated by sampling ten laser polarization vectors on vertices of a dodecahedron in half-space. The polarization vectors $\bm{l}$ used in this study are $(\pm1,\pm1,1)$, $(\pm\phi,0,1/\phi)$, $(\pm1/\phi,\phi,0)$, and $(0,\pm1/\phi,\phi)$, where $\phi = (1+\sqrt{5})/2$, these being the same vectors used within a previous work performed by Hein \textit{et al} \citep{Hein2012}. Armed with these laser polarization vectors one can then specify the transition dipole moment $\mu_j$ used in Eq. \eqref{eq:op_dipole} with $\mu_j = \bm{d}_j \cdot \bm{l}$, where $\bm{d}_j$ is the transition dipole pointing from the N$_\mathrm{B}$ to N$_\mathrm{D}$ atom of the $j$-th BChl.

\subsection{The scaled HEOM and high temperature approximation}

The correlation function of collective bath operator can be written as
\begin{equation}
\begin{split}
C_{j}\left(t\right) & = \frac{1}{\mathrm{Tr_B}\left\{e^{-\beta H_B}\right\}}\mathrm{Tr_B}\left\{e^{-\beta H_B}e^{iH_Bt/\hbar}\hat{B}_je^{-iH_Bt/\hbar}\hat{B}_j\right\} \\
       &=\frac{1}{\pi}\intop_{-\infty}^{\infty}d\omega D_{j}\left(\omega\right)\frac{e^{-i\omega t}}{1-e^{-\beta\hbar\omega}}\;, \label{eq:correlation}
\end{split}
\end{equation}
where the spectral distribution function is given by $D_{j}\left(\omega\right)=\hbar^{-1}\sum_{\xi}\left[c_{j\xi}^{2}/2m_{j\xi}\omega_{j\xi}\right]\delta\left(\omega-\omega_{j\xi}\right)$.
The spectral density of the bath is assumed to be same for each site, and the Drude model spectral density is employed in this study:
\begin{align}
D_{j}\left(\omega\right)=\frac{2\lambda}{\hbar}\frac{\omega\gamma}{\omega^{2}+\gamma^{2}}\;,\label{eq:Drude}
\end{align}
Where $\lambda = \lambda_j = \sum_{\xi}c_{j\xi}^{2}/2m_{j\xi}\omega_{j\xi}^2$ is the reorganization energy and $\gamma$ is the Drude decay constant. The correlation function now can be written as:
\begin{align}
C_{j}\left(t>0\right) & =\sum_{k=0}^{\infty}c_{k} e^{-v_{k}t}\;,\label{eq:correlationfunc}
\end{align}
where $v_{0}=\gamma$ and $v_{k}=2k\pi/\beta\hbar$ ($k\geqslant1$) are known as the Matsubara frequencies.
The constants $c_{k}$ are given by:
\begin{align}
c_{0}&=\frac{\lambda\gamma}{\hbar}\left[\cot\left(\frac{\beta\hbar\gamma}{2}\right)-i\right]\, \label{eq:c0}
\end{align} and
\begin{align}
c_{k}&=\frac{4\lambda\gamma}{\beta\hbar^2}\frac{v_{k}}{v_{k}^{2}-\gamma^{2}}\;\; ,\;\; \mathrm{for} \, k\geqslant1\;\;. \label{eq:ck}
\end{align}

Using the scaled HEOM approach developed by Shi and coworkers, \cite{Shi2009a} we apply the Ishizaki-Tanimura scheme \cite{Tanimura,Tanimura2} which truncates the hierarchy level at the $K^{th}$ Matsubara frequency by treating higher phonon frequencies ($v_{k>K}$) with the Markovian approximation: $v_{k}e^{-v_{k}t} \simeq \delta(t)$. The time evolution of the density operator can be described by the following set of hierarchically coupled equations of motion:

\begin{multline}
\frac{d}{dt}\hat{\rho}_{\boldsymbol{n}}=-\frac{i}{\hbar}\left[H_S,\;\hat{\rho}_{\boldsymbol{n}}\right]-\sum_{j=1}^{N}\sum_{k=0}^{K}n_{jk}v_{k}\hat{\rho}_{\boldsymbol{n}}-i\sum_{j=1}^{N}\sum_{k=0}^{K}\sqrt{\left(n_{jk}+1\right)\left|c_{k}\right|}\,\left[\hat{P}_j,\;\hat{\rho}_{\boldsymbol{n_{jk}^{+}}}\right]\\
-\sum_{j=1}^{N}\left(\sum_{m=K+1}^{\infty}\frac{c_{m}}{v_{m}}\right)\left[\hat{P}_j,\,\left[\hat{P}_j,\,\hat{\rho}_{\boldsymbol{n}}\right]\right]-i\sum_{j=1}^{N}\sum_{k=0}^{K}\sqrt{\frac{n_{jk}}{\left|c_{k}\right|}}\;\left(c_{k}\hat{P}_j\,\hat{\rho}_{\boldsymbol{n_{jk}^{-}}}-c_{k}^{*}\hat{\rho}_{\boldsymbol{n_{jk}^{-}}}\hat{P}_j\right)\;.\label{eq:HEOM}
\end{multline}
Within the above $\boldsymbol{n}$ is defined as one set of non-negative integers
$\boldsymbol{n}\equiv\{n_{1},n_{2},\cdots,n_{N}\}=\{\{n_{10},n_{11}\cdots,n_{1K}\},\cdots,\{n_{N0,}n_{N1}\cdots,n_{NK}\}\}$.
and $\boldsymbol{n_{jk}^{\pm}}$ refers to change the value of $n_{jk}$ to $n_{jk}\pm1$ in the global index $\boldsymbol{n}$. 
The density operator with all indices equal to $0$ is the system's reduced density operator (RDO) while all other
operators are the auxiliary density operators (ADOs). Here another truncation level $\mathcal{N}_{C}$ is introduced, defined as $\mathcal{N}_{C}=\sum_{j,k}n_{jk}$, which limits the total number of density operators to $C^{\mathcal{N}_{C}+(K+1)N}_ {\mathcal{N}_{C}}$. 

Unfortunately, simulating 2D electronic spectra by the HEOM method requires enormous computational resource. Chen et al.\ developed a modified version of HTA to HEOM by applying the Ishizaki-Tanimura truncating scheme to all the Matsubara frequencies \cite{Chen2010}. The modified HTA is capable of providing similar peak shapes and amplitude evolution to the original HEOM method, and the equations of motion can be described as:
\begin{multline}
\frac{d}{dt}\hat{\rho}_{\boldsymbol{n}}=-\frac{i}{\hbar}\left[H_S,\;\hat{\rho}_{\boldsymbol{n}}\right]-\gamma\sum_{j=1}^{N}n_{j}\hat{\rho}_{\boldsymbol{n}}-i\sum_{j=1}^{N}\,\left[\hat{P}_j,\;\hat{\rho}_{\boldsymbol{n_{j}^{+}}}\right]\\
-\sum_{j=1}^{N}\left(\frac{2\lambda}{\beta\hbar^2\gamma}-\frac{\lambda}{\hbar}\cot\left(\frac{\beta\hbar\gamma}{2}\right)\right)\left[\hat{P}_j,\,\left[\hat{P}_j,\,\hat{\rho}_{\boldsymbol{n}}\right]\right]-i\sum_{j=1}^{N}\;n_{j}\left(c_{0}\hat{P}_j\,\hat{\rho}_{\boldsymbol{n_{j}^{-}}}-c_{0}^{*}\hat{\rho}_{\boldsymbol{n_{j}^{-}}}\hat{P}_j\right)\;.\label{eq:mHTA}
\end{multline}

\newpage
 
\section{Results}
In all simulations the reorganization energy is set to be 35 cm$^{-1}$; this value has been employed in many previous works \citep{Chen2011}; The relaxation time constant, $\gamma^{-1}$, is assumed to be 100 fs$^{-1}$. Arcsinh scaling is applied to all 2D spectra by firstly, linear scaling the maximum signal $I_\mathrm{max} = 10$; the final signal is then obtained by calculating arcsinh($I$) $= \mathrm{ln}(I+\sqrt{1+I^2})$ \cite{Engel-2007}. To simulate the effect of the static disorder, a Gaussian distributed noise with a standard deviation of 25 cm$^{-1}$ is added to every site energy in the single-exciton Hamiltonian Eq.~\eqref{eq:H1ex}. 

\subsection{Exciton delocalization comparison between the \textit{apo}- and \textit{holo}- FMO}

\begin{figure}[h!]
\centering
\includegraphics[scale=0.60]{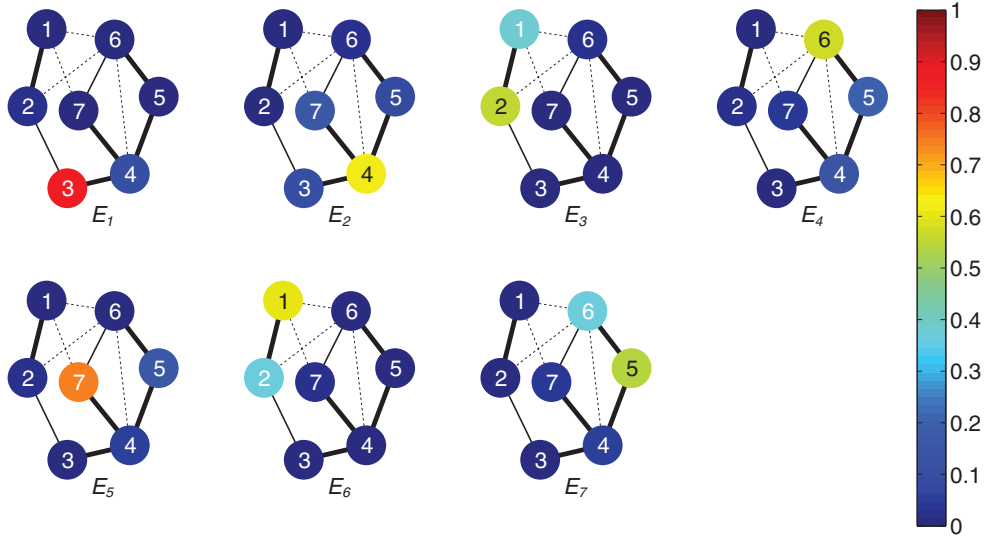}
\caption{\label{fig:FMO7ex} (Color) The probability density distribution of excitonic states on each BChl in \textit{apo}-FMO. The thick lines represent the excitonic couplings that are greater than 50 \% of the largest excitonic coupling. The thin lines and the dotted lines represent the couplings that are 30-50 \% and 10-30 \% of the largest coupling respectively.}
\end{figure}

\begin{figure}[h!]
\centering
\includegraphics[scale=0.60]{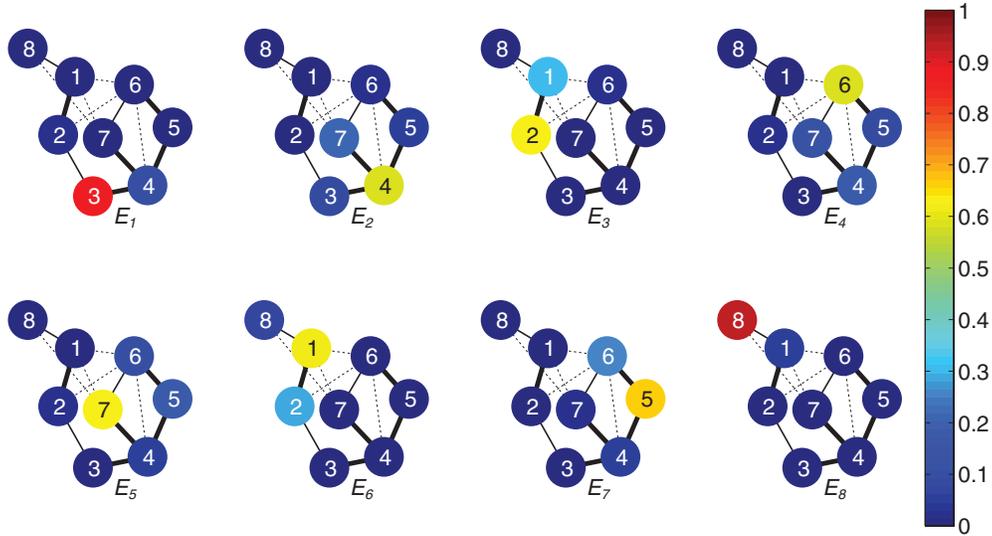}
\caption{\label{fig:FMO8ex} (Color) The probability density distribution of excitonic states on each BChl in \textit{holo}-FMO. The thick lines represent the excitonic couplings that are greater than 50 \% of the largest excitonic coupling. The thin lines and the dotted lines represent the couplings that are 30-50 \% and 10-30 \% of the largest coupling respectively.}
\end{figure}

Because the exciton energy transfer pathway is highly depending on the spacial overlap between excitonic wave functions on each site, a comparison between the \textit{apo}- and \textit{holo}-FMO delocalized excitons has been completed by calculating the site (BChl) occupation probabilities of each exciton. These probability are obtained by solving the eigenvectors of the single-exciton Hamiltonian Eq.~\eqref{eq:H1ex} from each form. The probability density distribution of each excitonic state of \textit{apo}- and \textit{holo}-FMO are shown in Fig.~\ref{fig:FMO7ex} and Fig.~\ref{fig:FMO8ex}, respectively. The calculated values for the probability density for all excitons are provided in Tables ~\ref{tab:FMO7exdel} and ~\ref{tab:FMO8exdel}. 

\begin{table}[h!]
\caption{Squares of the eigenvector elements of \textit{apo}-FMO} \label{tab:FMO7exdel}
\begin{center}
\begin{tabular}{c c c c c c c c}
 \hline\hline
 & $E_1$ & $E_2$ & $E_3$ & $E_4$ & $E_5$ & $E_6$ & $E_7$\\
 \hline
$S_1$ & 0.00098 &0.00936 &	\textbf{0.37792} & 0.00182 & 0.00371 & \textbf{0.60617} & 0.00004 \\
$S_2$ & 0.01660 &0.00484 &	\textbf{0.56177} & 0.02734 & 0.02476 & \textbf{0.36325} & 0.00145 \\
$S_3$ & \textbf{0.87520} &0.09882 & 0.00786 & 0.00936 & 0.00457 &	0.00372 & 0.00047 \\
$S_4$ & \textbf{0.10106} &\textbf{0.61940} & 0.00977 & \textbf{0.15324} & 0.05440 &	0.00003 & 0.06212 \\
$S_5$ & 0.00243 &0.08566 &	0.01492 & \textbf{0.19006} & \textbf{0.16516} &	0.00516 & \textbf{0.53661} \\
$S_6$ & 0.00081 &0.02144 &	0.02775 & \textbf{0.57384} & 0.01061 &	0.00581 & \textbf{0.35975} \\
$S_7$ & 0.00292 & \textbf{0.16049} & 0.00002 &0.04435 & \textbf{0.73681} &	0.01587 & 0.03956 \\
\hline\hline
\end{tabular}\par
\bigskip
Site population coefficients for each excitonic state for the 7-site system. $E_n$/$S_m$: The state of the $n^{th}$ exciton / $m^{th}$ site in site basis; should be read as, the degree to which the $n^{th}$ exciton is present on the $m^{th}$ site.  The sites with larger populations in each exciton are shown in bold.
\end{center}
\end{table}

\begin{table}[h!]
\caption{Squares of the eigenvector elements of \textit{holo}-FMO} \label{tab:FMO8exdel}
\begin{center}
\begin{tabular}{c c c c c c c c c}
 \hline\hline
 & $E_1$ & $E_2$ & $E_3$ & $E_4$ & $E_5$ & $E_6$ & $E_7$ & $E_8$\\
 \hline
$S_1$ & 0.00094 & 0.00990 & \textbf{0.30699} & 0.00104 & 0.00056 & \textbf{0.62309} & 0.00038 & 0.05709 \\
$S_2$ & 0.01794 & 0.00787 & \textbf{0.62873} & 0.01932 & 0.01984 & \textbf{0.29432} & 0.00153 & 0.01044 \\
$S_3$ & \textbf{0.87897} & 0.09301 & 0.00836 & 0.01196 & 0.00402 & 0.00314 & 0.00043 & 0.00012 \\
$S_4$ & 0.09624 & \textbf{0.59312} & 0.01395 & \textbf{0.17855} & 0.06248 & 0 & 0.05558 & 0.00006 \\
$S_5$ & 0.00187 & 0.06057 & 0.01128 & 0.09129 & \textbf{0.17284} & 0.00053 & \textbf{0.66160} & 0.00001 \\
$S_6$ & 0.00093 & 0.02093 & 0.02835 & \textbf{0.57981} & 0.09980 & 0.00501 & \textbf{0.26023} & 0.00494 \\
$S_7$ & 0.00312 & \textbf{0.21459} & 0.00064 & \textbf{0.11751} & \textbf{0.63738} & 0.00418 & 0.01845 & 0.00414 \\
$S_8$ & 0 & 0.00001 & 0.00170 & 0.00051 & 0.00307 & 0.06973 & 0.00179 & \textbf{0.92318} \\
\hline\hline
\end{tabular}\par
\bigskip
Site population coefficients for each excitonic state for the 8-site system. $E_n$/$S_m$: The state of the $n^{th}$ exciton / $m^{th}$ site in site basis; should be read as, the degree to which the $n^{th}$ exciton is present on the $m^{th}$ site.  The sites with larger populations in each exciton are shown in bold.
\end{center}
\end{table}

In the excitonic wave function overlap of \textit{apo}-FMO, two major EET pathways can be identified. The first one is through excitons 6 and 3 that are both delocalized over BChl 1 and 2. The energy will then be funnelled downstream to exciton 1, which is mostly localized on the energy sink of FMO complex--BChl 3. The second pathway features the excitonic wave packet starting from exciton 7, flowing through excitons 5 and/or 4, then passing through exciton 2 and finally to exciton 1. This result is similar to previous studies using another green sulfur bacteria species, \textit{Chlorobium tepidum} \citep{Cho2005}. The only major difference in this result is the starting exciton number from both pathways. In the work of Cho \textit{et al.} \citep{Cho2005}, the two pathways they obtained are: 7 $\rightarrow$ 3 $\rightarrow$ 1 and 6 $\rightarrow$ 5,4 $\rightarrow$ 2 $\rightarrow$ 1, where in our case we find: 6 $\rightarrow$ 3 $\rightarrow$ 1 and 7 $\rightarrow$ 5,4 $\rightarrow$ 2 $\rightarrow$ 1.

For the \textit{holo}-FMO case, the BChl 8 modifies the EET in two ways. Due to the vicinity of BChl 8 to BChl 1, it leads directly  to an analoguous path to the first pathway discussed in the previous paragraph for the \textit{apo}-FMO system: 8 $\rightarrow$ 6 $\rightarrow$ 3 $\rightarrow$ 1. Furthermore, the second pathway is similarly affected by the enhanced delocalization of excitons 4 and 5 over BChl 4, 5, 6, and 7 and thus the excitonic wave function overlap between excitons 4 and 5 is increased. This larger spacial overlap may facilitate the energy transfer through the second pathway: 7 $\rightarrow$ 5,4 $\rightarrow$ 2 $\rightarrow$ 1.

\subsection{2D electronic spectra without static disorder}

\begin{figure}[h!]
\centering
\includegraphics[scale=0.6]{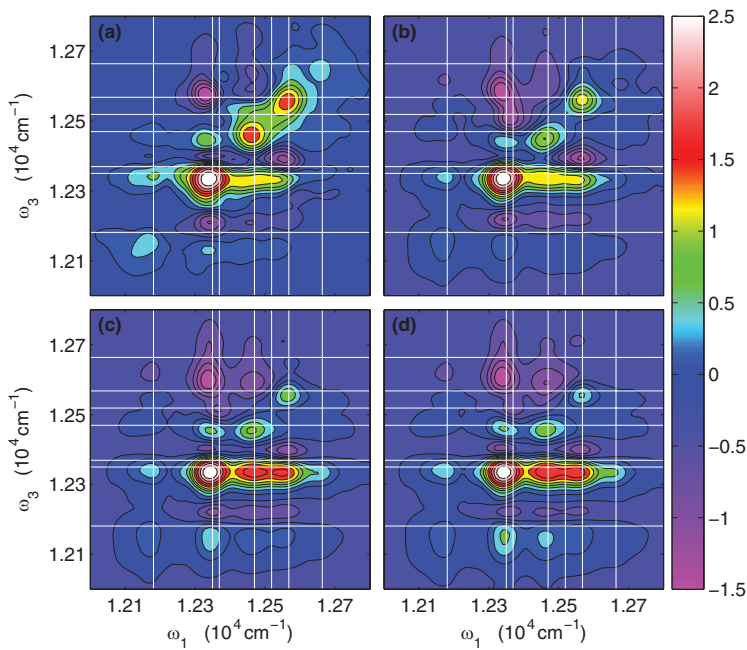}
\caption{\label{fig:FMO7_77K_noSD} (Color) Simulated 2D electronic spectrum of \textit{apo}-FMO complex from \textit{P. aestuarii} at 77 K without static disorder; a waiting time, $t_2$, of: (a) 0 fs, (b) 192 fs, (c) 384 fs, and (d) 1024 fs was used.}
\end{figure}

\begin{figure}[h!]
\centering
\includegraphics[scale=0.6]{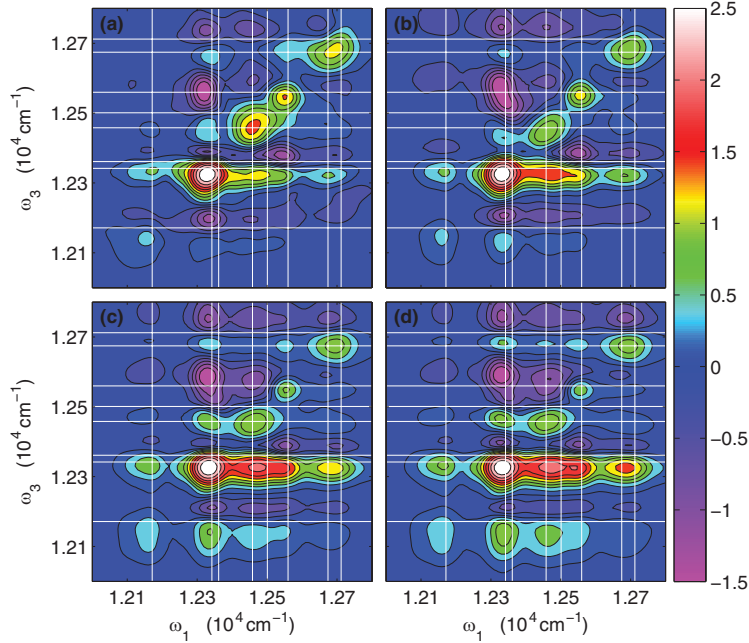}
\caption{\label{fig:77K_noSD} (Color) Simulated 2D electronic spectrum of \textit{holo}-FMO complex from \textit{P. aestuarii} at 77 K without static disorder; a waiting time, $t_2$, of: (a) 0 fs, (b) 192 fs, (c) 384 fs, and (d) 1024 fs was used.}
\end{figure}

Under the condition that static disorder is not included, we have simulated multiple 2D electronic spectra with different waiting times, $t_2$, from 0 fs to 1024 fs, using an interval of 32 fs. The 2D photon-echo spectra of \textit{apo}- and \textit{holo}-FMO at 77 K are shown in Figs.~\ref{fig:FMO7_77K_noSD} and \ref{fig:77K_noSD}, respectively. The white lines represent the eigenenergies of the single-exciton Hamiltonian. It is worth noting here that for the Hamiltonian used in this study Eq.~\eqref{eq:H1ex}, the eigenenergies of excitons 2 and 3 nearly degenerate to each other on our global scale in both forms of FMO; in \textit{apo}, \textit{holo} form the energy differences are 19.33 cm$^{-1}$, 19.36 cm$^{-1}$, respectively. 

Fig.~\ref{fig:FMO7_77K_noSD} shows a simulated 2D electronic spectrum of the \textit{apo}-FMO, in which four diagonal peaks can be consistently identified throughout the simulation, these being excitons 1, 2-3, 4, and 6. The exciton 7 diagonal peak is only distinguishable within the first 300 fs of simulation. Along the time evolution, the EET from excitons 4,6 $\rightarrow$ 2 becomes increasingly significant; the same trend is also found in the exciton 7 $\rightarrow$ 3 transfer. Compared with the \textit{apo} form, the \textit{holo} form shown in Fig.~\ref{fig:77K_noSD} has a long-lived diagonal peak at exciton 7-8 region. The 2D spectra data compared with the previous excitonic state delocalization analysis informs that the \textit{apo}-FMO spectrum shows a broader peak around exciton 4 which implies an enhanced transition from excitons 5 and 4. Additionally, at the lower diagonal there exists a significant cross-peak between exciton 7-8 to 2-3, which is very likely to be contributed to by the energy transfer from exciton 8 to 3. There are also upper diagonal cross-peaks between excitons 8, 6, and 3 which is consistent with the first pathway of EET (8 $\rightarrow$ 6 $\rightarrow$ 3 $\rightarrow$ 1) in our previous analysis. 

%

\begin{figure}[h!]
\centering
\includegraphics[scale=0.6]{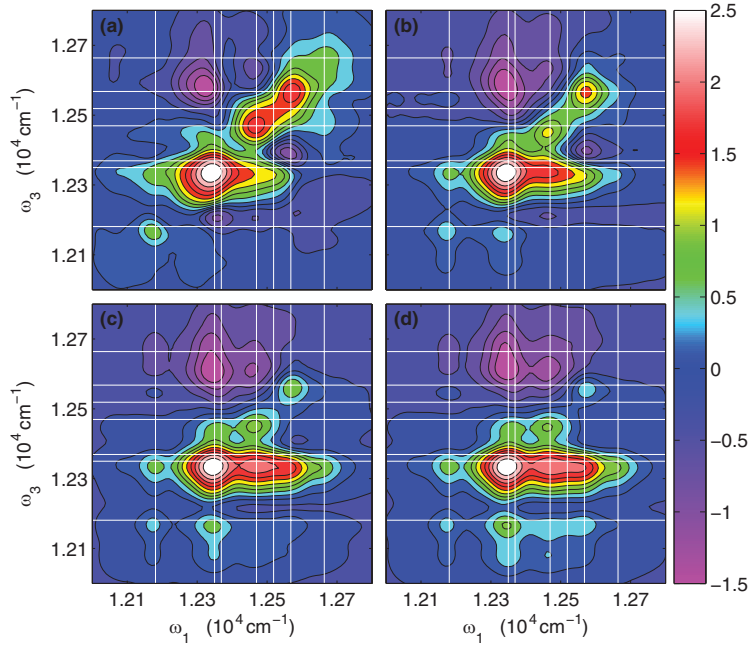}
\caption{\label{fig:FMO7_150K_noSD} (Color) Simulated 2D electronic spectrum of \textit{apo}-FMO complex from \textit{P. aestuarii} at 150 K without static disorder; a waiting time, $t_2$, of: (a) 0 fs, (b) 192 fs, (c) 384 fs, and (d) 1024 fs was used.}
\end{figure}

\begin{figure}[h!]
\centering
\includegraphics[scale=0.6]{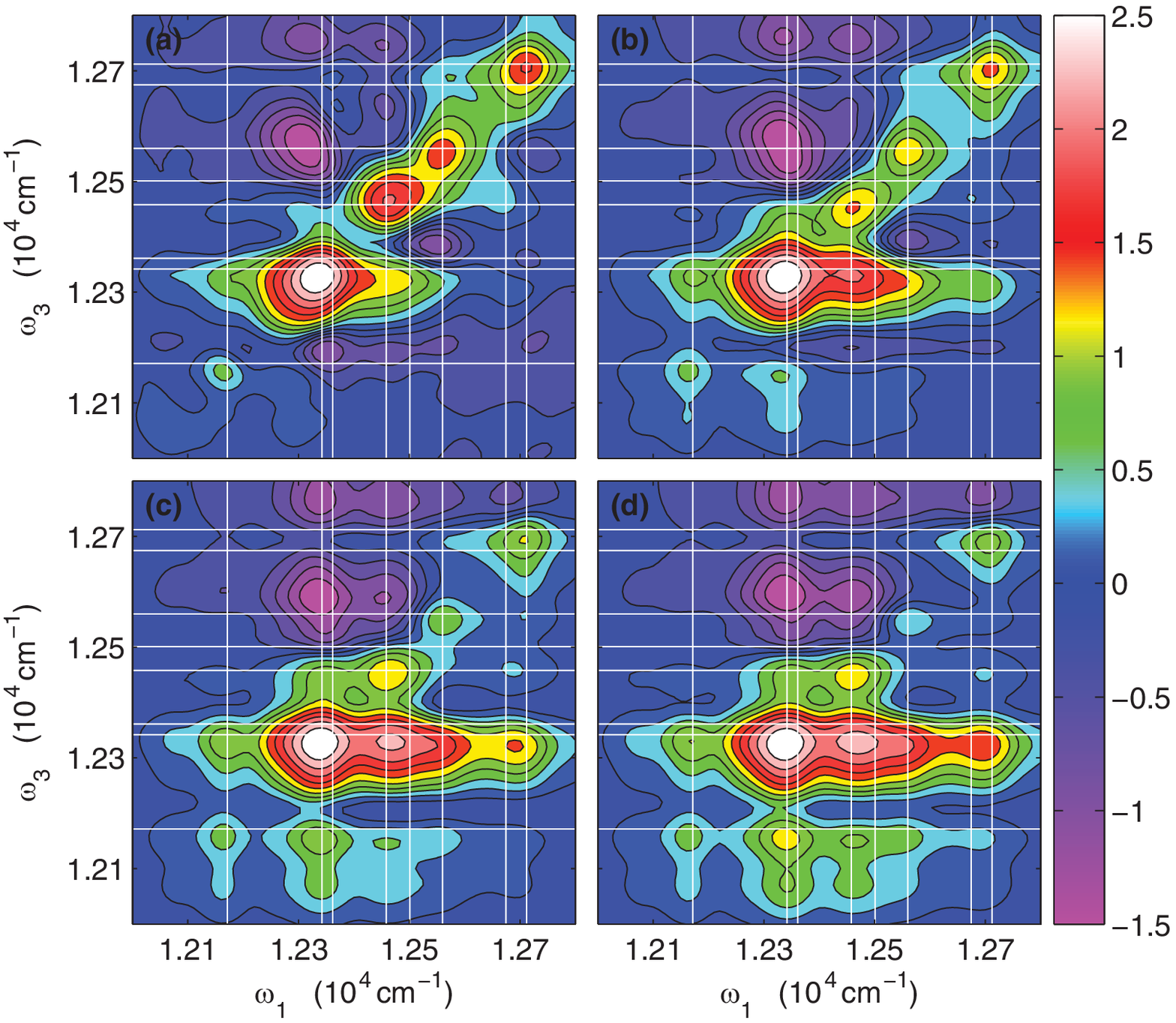}
\caption{\label{fig:150K_noSD} (Color) Simulated 2D electronic spectrum of \textit{holo}-FMO complex from \textit{P. aestuarii} at 150 K without static disorder; a waiting time, $t_2$, of: (a) 0 fs, (b) 192 fs, (c) 384 fs, and (d) 1024 fs was used.}
\end{figure}

To investigate temperature effects on the EET, the 2D spectra of both forms of FMO have also been simulated at 150 K; the results are shown in Figs.~\ref{fig:FMO7_150K_noSD} and \ref{fig:150K_noSD}. The homogeneous and inhomogeneous broadening significantly increases compared to the 77 K spectrum. The enhanced EET from exciton 5 to 4 in the \textit{holo}-FMO is still very apparent at 150 K, and a similar trend exists for the exciton 8 $\rightarrow$ 6 $\rightarrow$ 3 transfer can be also seen from the lower diagonal of the same spectrum. However, it is worth noticing that the upper diagonal part of the 8 $\rightarrow$ 6 $\rightarrow$ 3 transfer no longer exists at higher temperatures when compared to 77 K spectrum of the \textit{holo}-FMO. 

\begin{figure}[h!]
\centering
\includegraphics[scale=0.6]{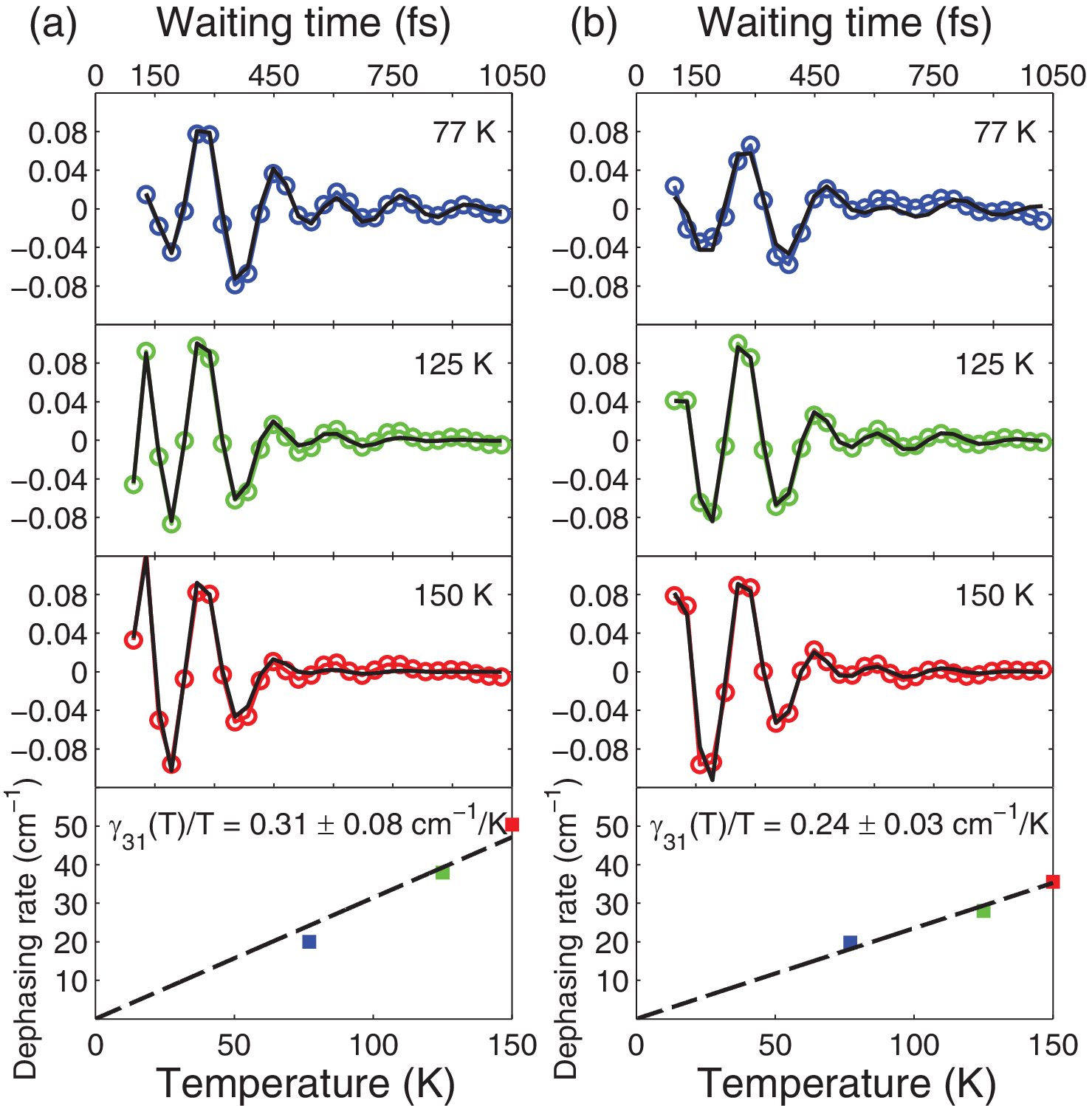}
\caption{\label{fig:FMO_fit} (Color) Temperature dependence of dephasing rate from (a) \textit{apo}-FMO, and (b) \textit{holo}-FMO. The beating signal is obtained from integrated cross-peak amplitude after subtraction of population transfer modeled by decaying exponential function from 77K, 125 K, and 150 K (colored lines with circles). The sum of two exponentially decaying functions we used for fitting the beating signal are shown in black solid lines. The dephasing rates for the 3 $\rightarrow$ 1 transition $\gamma_{31}$ and its temperature dependence, fitted linearly, is shown at the bottom (dashed black line).}
\end{figure}

The dephasing rate of the excitonic transfer 2,3 $\rightarrow$ 1 are analysed by fitting the oscillation of the corresponding cross-peak (lower diagonal) from the 2D spectra. The analysis has been done by first integrating the cross-peak amplitude over a 20 cm$^{-1} \times$ 20 cm$^{-1}$ area, and then apply a multi-exponential fit to the signal in an effort to model the population transfer. After subtracting the population transfer part of the signal, the remainder is fitted by the sum of two exponentially decaying sinusoidal functions (Fig. ~\ref{fig:FMO_fit}). Simulating results with waiting time $T$ less than 96 fs are not included for avoiding the pulse overlap effects in experimental studies \cite{Engel2}. We have found that the dephasing rates all increase linearly with the raising temperature (77K, 125K, and 150K are the temperature used in this study). For the \textit{apo}-FMO the temperature dependence of the dephasing rates are $\gamma_{21}(T)/T = 0.33 \pm 0.04$ and $\gamma_{31}(T)/T = 0.31 \pm 0.08$, whereas in \textit{holo}-FMO the numbers are $\gamma_{21}(T)/T = 0.23 \pm 0.02$ and $\gamma_{31}(T)/T = 0.24 \pm 0.03$. This linear dependence of dephasing rate on temperature is similar to the experimental results in Ref.~\citep{Engel2} the difference being that our numbers are smaller. This suggests that in the physical condition the relaxation time constant, $\gamma^{-1}$, should be smaller than 100 fs$^{-1}$.

\subsection{2D electronic spectra with static disorder}

\begin{figure}[h!]
\centering
\includegraphics[scale=0.6]{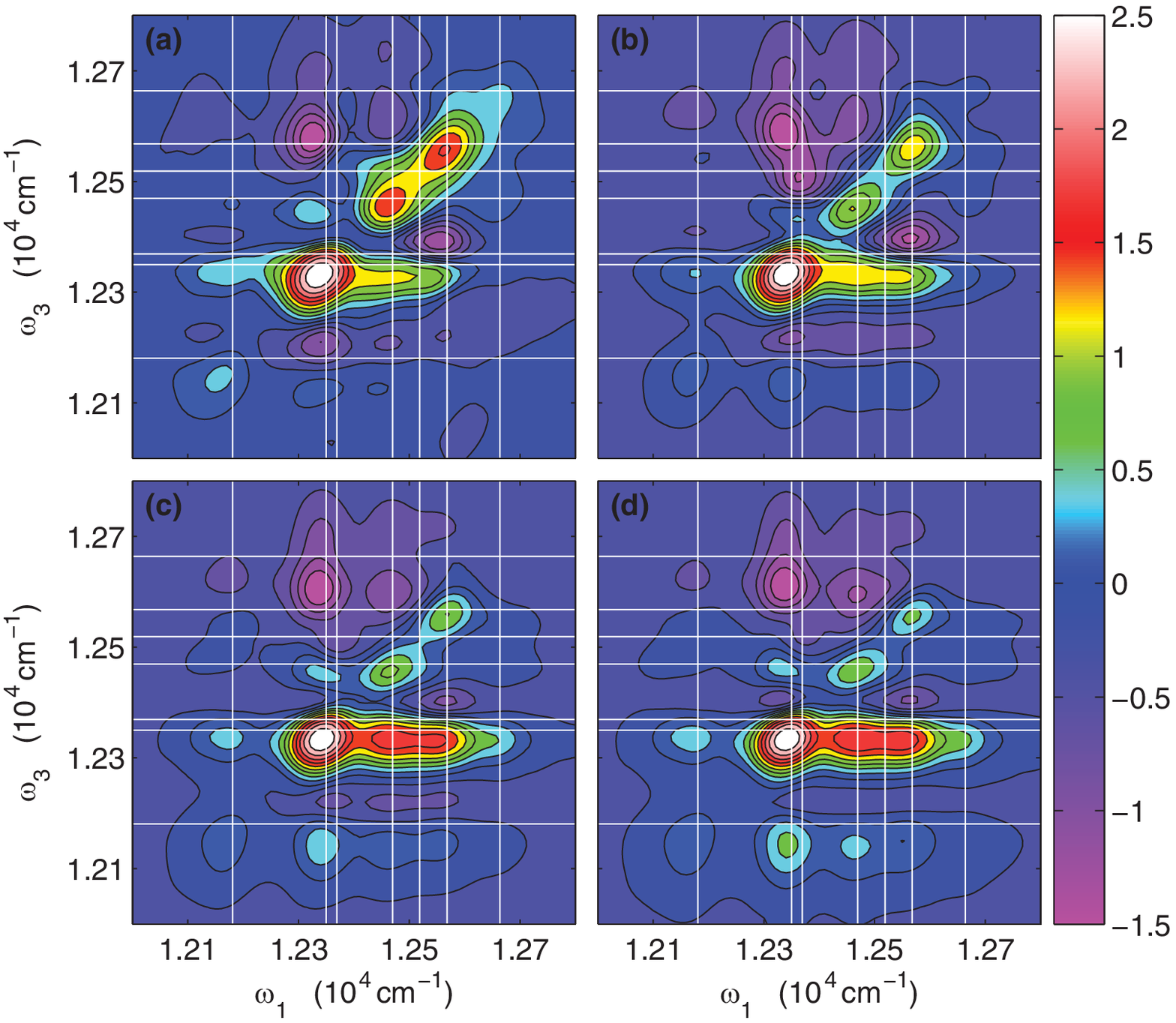}
\caption{\label{fig:FMO7_77K_SD25} (Color) Simulated 2D electronic spectrum of \textit{apo}-FMO complex from \textit{P. aestuarii} at 77 K with static disorder of 25 cm$^{-1}$. The waiting time, $t_2$, are taken at: (a) 0 fs, (b) 192 fs, (c) 384 fs, and (d) 1024 fs.}
\end{figure}

\begin{figure}[h!]
\centering
\includegraphics[scale=0.6]{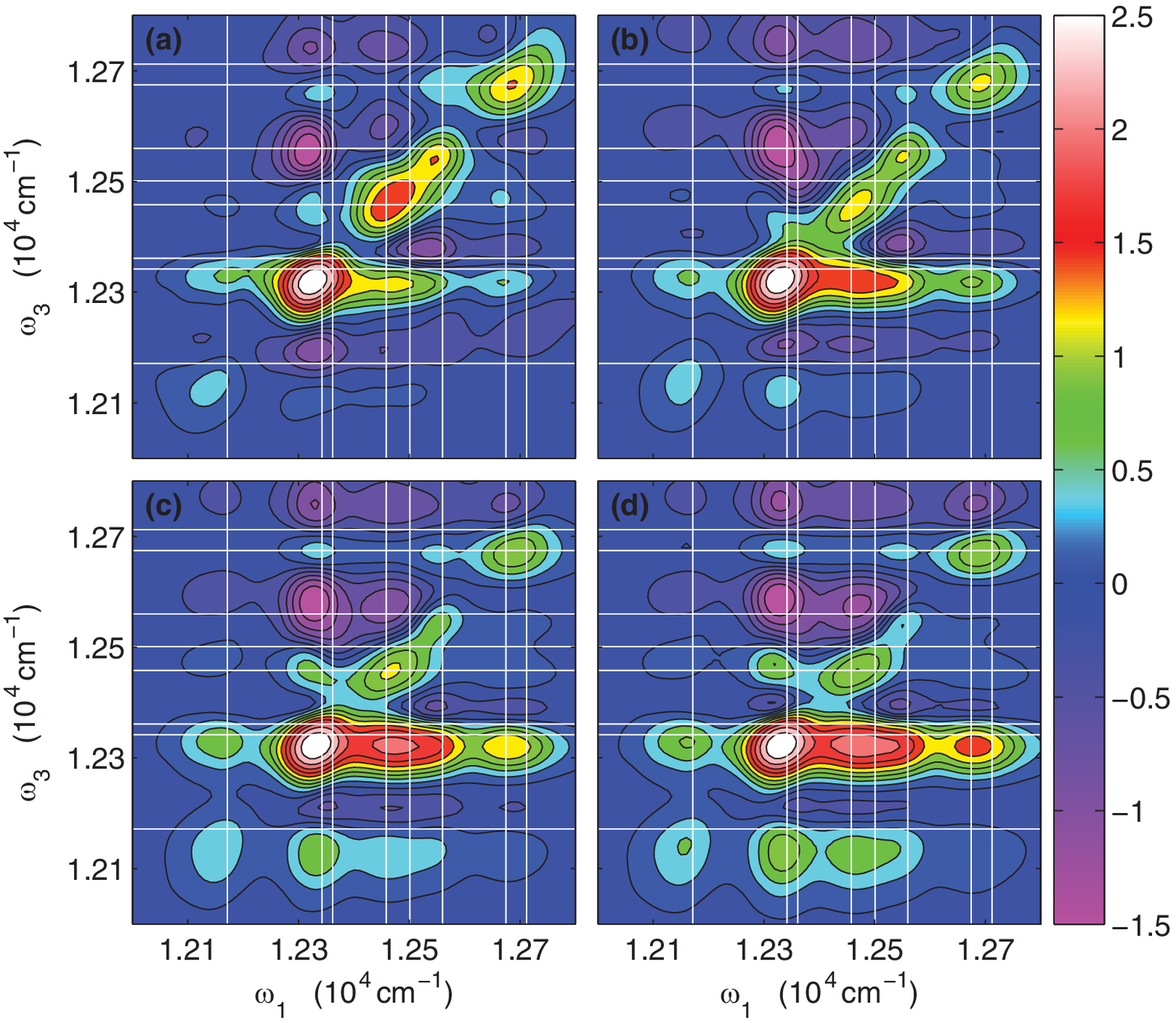}
\caption{\label{fig:77K_SD25} (Color) Simulated 2D electronic spectrum of \textit{holo}-FMO complex from \textit{P. aestuarii} at 77 K with static disorder of 25 cm$^{-1}$. The waiting time, $t_2$, are taken at: (a) 0 fs, (b) 192 fs, (c) 384 fs, and (d) 1024 fs.}
\end{figure}

To investigate how static disorder affects the EET, a Gaussian distributed noise with standard deviation of 25 cm$^{-1}$ is applied on every site energy, $\epsilon_j$, of Eq.~\eqref{eq:H1ex}. The simulated spectrum are averaged over 1000 samples of different laser polarizations along with static disorder. Figs.~\ref{fig:FMO7_77K_SD25} and \ref{fig:77K_SD25} show the result of \textit{apo}- and \textit{holo}-FMO, respectively, at 77 K with different waiting time as described in the captions. 

The spectroscopic features in Figs.~\ref{fig:FMO7_77K_SD25} and \ref{fig:77K_SD25} are very similar to those features within the 2D spectra calculated without static disorder (Figs.~\ref{fig:FMO7_77K_noSD} and \ref{fig:77K_noSD}) except for the increased homogeneous and inhomogeneous broadening; also for the upper diagonal of Fig.~\ref{fig:77K_SD25} the cross-peaks of excitons (8,6) and (8,3) still exist, implying that the disappearance of these cross-peaks in Fig.~\ref{fig:150K_noSD} may not merely be due to an increase in static disorder. 

%

\section{Conclusions}
In this study we generated the 2D photon-echo electronic spectra of \textit{apo}- and \textit{holo}-FMO at different temperatures by using a high temperature approximation of HEOM. We found that the 8th BChl can affect the EET through both available pathways. From our simulated 2D spectrum, the cross-peak of exciton (8,3) shows that the 8th BChl is able to enhance the EET through the 8 $\rightarrow$ 6 $\rightarrow$ 3 $\rightarrow$ 1 pathway due to an enhancement in the coupling, although the excitonic wave function overlap between exciton 3 and 1 is quite low and the transfer rate is comparatively slower than the other pathway. Along the 7 $\rightarrow$ 5,4 $\rightarrow$ 2 $\rightarrow$ 1 pathway, the 8th BChl increases the excitonic wave function overlap between exciton 4 and 5 and hence facilitates the downward energy transfer. Introducing static order into the system increases both the levels of homogeneous and inhomogeneous broadening; yet, this static interference did not affect the existence of the upper diagonal cross-peaks of exciton (8,6) and (8,3), suggesting these peaks are more sensitive to temperature than to static disorder. 

In both forms of the FMO complex we observed cross-peak oscillation which last at minimum $\sim$600 fs. To compare \textit{apo} and \textit{holo}-FMO, we analysed the dephasing rate of the important 3 $\rightarrow$ 1 excitonic transition and found that the 8th BChl decreases the dephasing rate slightly. This reduction in dephasing alleviates excitonic energy loss to the environment. Furthermore, we have also shown that this rate is linearly dependent to the temperature, which is in agreement with findings in other experimental and theoretical studies \citep{Engel2, Hein2012}.

\section{Acknowledgement}
The authors would like to thank the ITaP Research Computing of Purdue University and support from National Science Foundation Centers for Chemical Innovation: CHE-1037992.

\end{document}